\documentclass[amsmath,amssymb,prl,twocolumn]{revtex4}                                                                           
\usepackage{graphicx}
\usepackage{color}
\usepackage{bm}
\usepackage{longtable}                                
\usepackage{graphics}
\date{\today}
\begin{document}
\title{Superdense coding using the quantum superposition principle}

\author{Dipankar Home}
\altaffiliation{dhome@bosemain.boseinst.ac.in}
\affiliation{CAPSS, Dept. of Physics, Bose Institute, Salt Lake, 
Kolkata-700091, India}

\author{A. K. Pan}
\altaffiliation{apan@bosemain.boseinst.ac.in}
\affiliation{CAPSS, Dept. of Physics, Bose Institute, Salt Lake, 
Kolkata-700091, India}

\author{S. Adhikari}
\altaffiliation{satyabrata@bose.res.in}
\affiliation{S. N. Bose National Centre for Basic Sciences,
Salt Lake, Kolkata 700 098, India}

\author{A. S. Majumdar}
\altaffiliation{archan@bose.res.in (corresponding author)}
\affiliation{S. N. Bose National Centre for Basic Sciences,
Salt Lake, Kolkata 700 098, India}

\author{M. A. B. Whitaker}
\altaffiliation{a.whitaker.qub.ac.uk}
\affiliation{Physics Department, Queen's University, Belfast BT7 1NN,
Northern Ireland}

\begin{abstract}
By sending a classical two-level system, one can  transfer information about 
only \emph{two} distinguishable outcomes. Here we show that in quantum 
mechanics, using  both the spin and path degrees of freedom of a spin-$1/2$ 
particle, and a Mach-Zehnder type interferometric arrangement with two suitable 
Stern-Gerlach detectors, it is possible to transfer information about 
\emph{four} distinguishable outcomes. This procedure does \emph{not} require 
using quantum entanglement as a resource as in the well-known protocol of
dense coding, but instead hinges entirely on the 
quantum superposition principle. We also study probabilistic dense coding
using our set-up and show that the dense coding scheme using quantum 
superposition cannot be optimized any further
by extending the interferometric arrangement with more beam splitters.
\end{abstract}

\maketitle

\emph{Introduction}

The quantum mechanical description of the nature of physical systems is 
enigmatic, and continues to spring surprises. On the one hand, it baffles 
specialists with its interpretational puzzles
to do with measurement \cite{meas}, nonlocality \cite{nonloc} and information
\cite{info}. On the other hand, it 
fascinates all with the unfolding of novel features associated with the 
superposition principle, as well as the applications of quantum entanglement.
Quantum superposition is responsible for several fascinating phenomena
 such as Bose-Einstein condensation \cite{bec} 
and superfluidity \cite{superfl}. 
 Entanglement is the
key ingredient for information processing protocols such as error 
correction \cite{qcomp}  and teleportation \cite {teleport}. 

Much is advocated about the utility of quantum entanglement as
resource for performing tasks impossible through classical means, and quite
justly so. But it is rather important to demarcate the useful tasks that may 
indeed be possible with the aid of the linear structure of quantum mechanics
embossed by the superposition principle without 
taking recourse to the entanglement of two or more particles. In this context
it is worthwhile to recall the debate on the necessity of using entangled
pairs of particles for key generation in quantum cryptography 
\cite{bennett,ekert}. Such distinction of the applications of purely quantum
superposition,
apart from its interesting pedagogical aspects, could be of operational 
significance too, since it might be considerably harder to set up and maintain 
entanglement at the practical level. 

The transfer of information in quantum theory through superdense coding 
\cite {dens} is regarded as one of the
major applications of quantum entanglement. In contrast,
in this work we demonstrate the possibility
of superdense coding employing the superposition principle for a single
particle using its path and spin degrees of freedom. This indeed is an 
example of a surprising application of quantum mechanics, since it is 
contrary to the widely 
subscribed notion of the entanglement of two particles being essential for 
dense coding \cite{dens,mermin}.

Classically, by sending a two-level system one can transfer information about 
only two distinguishable outcomes. In other words, only one bit may be encoded 
in one spin-1/2 particle. However, it is well-known that by the use of 
entanglement, the technique of dense coding \cite{dens} is able to transfer 
information about four distinguishable outcomes by sending a two-level system
that is prior entangled with another two-level system with the receiver. 
In this technique, Alice and Bob 
share the two members of an EPR pair of states. Alice then codes the required 
information into the spin-state of her member of the pair, and then sends this 
particle to Bob who carries out a Bell-basis measurement to obtain the 
information. The technique demonstrates that shared entanglement can enable 
Alice and Bob to enhance the capacity of a shared quantum channel 
up to the Holevo limit \cite{jaegar} by transmitting two bits of information 
using two qubits. Mattle et al \cite{mattle} have put the technique into 
practice using polarisation-entangled photons, and other important work on 
dense coding includes a fundamental discussion of Mermin \cite{mermin}, 
a dense coding 
protocol for continuous variables due to Braunstein and 
Kimble \cite{braunstein}, and 
achievement of this scheme by Li et al \cite{li}. Full experimental distinction 
between all four Bell states is still an ongoing task \cite{expt}. 

It is therefore interesting to explore the viability of an alternative scheme 
for transferring information about four distinguishable outcomes which does not
require quantum entanglement as a resource, but rather relies on the quantum 
superposition principle. The scheme presented in this paper utilises both the 
spin and path degrees of freedom of a spin-$1/2$ particle. 
The discussion of the 
scheme is in terms of spin-$1/2$ particles, such as neutrons, 
and uses a variant 
of the Mach-Zehnder interferometer, with suitable manipulations of both the 
spin and path degrees of freedom.  However the scheme works equally well for 
photons with appropriate polarising and analysing devices.

\emph{The setup}

In the variant of the Mach-Zender interferometer we are using here (Fig.1), 
an input spin-$1/2$ particle with an initial spin polarised state 
$\left|\uparrow\right\rangle_{z}$ is first passed through a spin rotator (SR) 
(in which a uniform magnetic field is directed along the $\widehat{x}$-axis) 
before  it is incident on the first beam splitter (BS1) of this setup (Fig.1). 
The action of SR is to change the initial spin state $\left|\uparrow\right\rangle_{z}=\frac{1}{\sqrt{2}}\left(|\rightarrow\rangle_{x}+ |\leftarrow\rangle_{x}\right)$ to the state, say,  $\left|\chi\right\rangle$  given by
\begin{equation}
\left|\chi\right\rangle=\frac{1}{\sqrt{2}}\left(|\rightarrow\rangle_{x}+ e^{i\delta} |\leftarrow\rangle_{x}\right)
\end{equation}
where $\delta$ is the relative phase shift between $|\rightarrow\rangle_{x}$ 
and $|\leftarrow\rangle_{x}$ introduced by SR.
In passing through BS1 with both the reflection and transmission probabilities 
$1/2$, the input particle can emerge along either the transmitted or the 
reflected channel. The state of the emergent particle in  either of these 
channels corresponds respectively to either $\left|\psi_{1}\right\rangle$ 
or $\left|\psi_{2}\right\rangle$ which have a relative phase shift of $(\pi/2)$
between them arising because of the reflection from BS1. 
Note that, $\left|\psi_{1}\right\rangle$ and $\left|\psi_{2}\right\rangle$ are 
 eigenstates of the projections operators $P(\psi_{1})$ and $P(\psi_{2})$ 
respectively, which pertain to  measurements determining \textit{`which 
channel'} the particle is in. For example, the results of such  measurements 
for the transmitted (reflected) channel with binary alternatives are given by 
the eigenvalues of $P(\psi_{1})$ ($P(\psi_{2})$); the eigenvalue $+1$ $(0)$ 
corresponds to the particle being found (not found) in the channel represented 
by $\left|\psi_{1}\right\rangle$($\left|\psi_{2}\right\rangle$). 

Next, a `path' phase shifter (PS) is applied along one of the channels, say  
$\left|\psi_{2}\right\rangle$, that introduces a relative phase shift, say 
$\phi$, between the states  $\left|\psi_{1}\right\rangle$ and 
$\left|\psi_{2}\right\rangle$. Reflections from the two mirrors M1 and M2 do 
not lead to any net relative phase shift between the states  
$\left|\psi_{1}\right\rangle$ and $\left|\psi_{2}\right\rangle$.

After all the above mentioned operations, the total state is given by the
spin state and the superposition of the two path states given by
\begin{eqnarray}
\left|\Psi\right\rangle_{SR+PS}=\frac{1}{\sqrt{2}}\left(\left|\psi_{1}\right\rangle +i e^{i\phi}\left|\psi_{2}\right\rangle\right)\left|\chi\right\rangle
\end{eqnarray}


\begin{figure}[h]
{\rotatebox{0}{\resizebox{9.0cm}{7.5cm}{\includegraphics{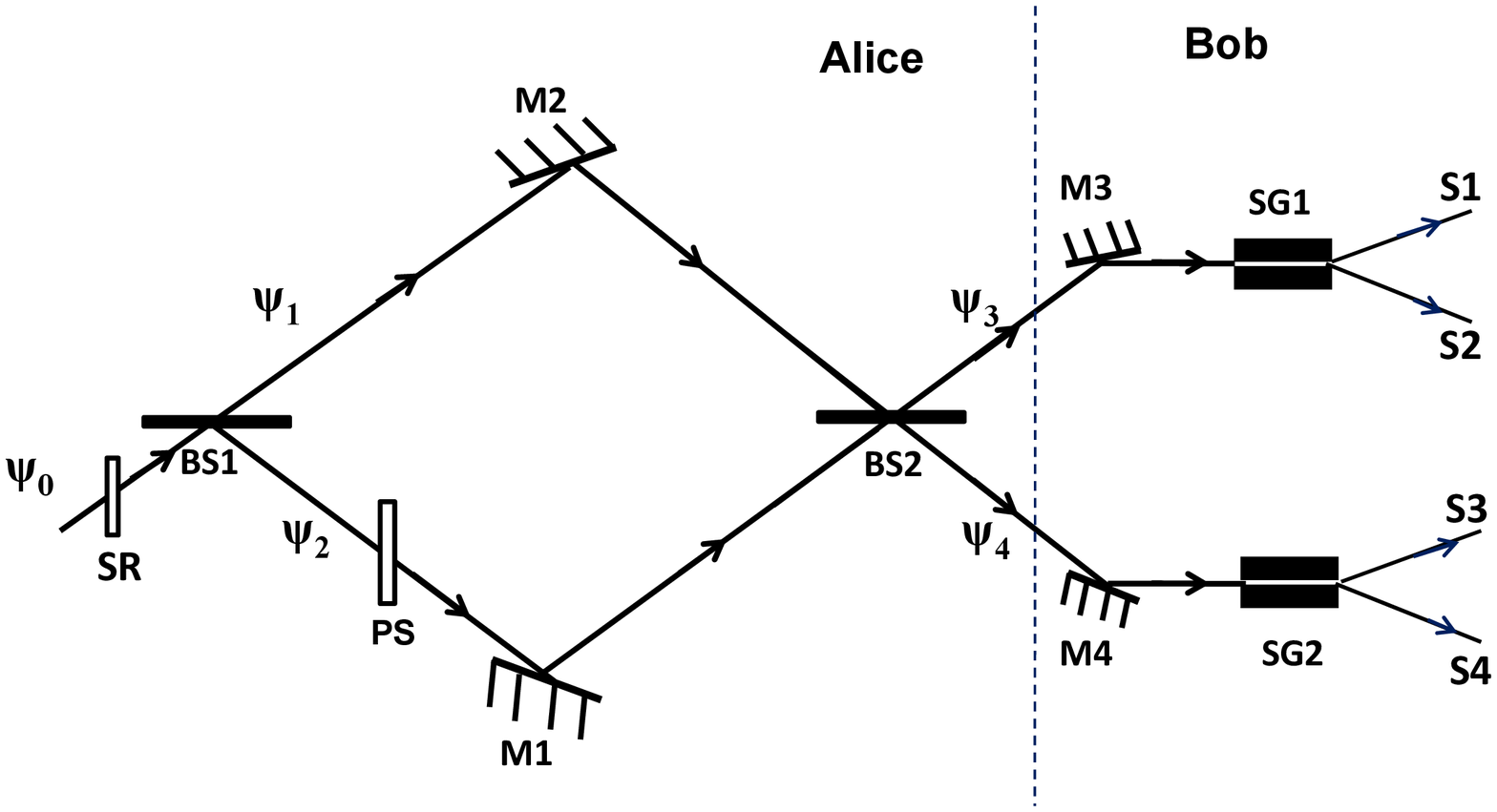}}}}

\caption{\footnotesize A spin-$1/2$ particle (say, a neutron) with an initial 
spin polarised state $\left|\uparrow\right\rangle_{z}$ is first passed through 
a spin rotator (SR)  before entering this Mach-Zehnder type setup through a 
beam splitter BS1. A phase shifter (PS) is placed along the channel 
$|\psi_{2}\rangle$. The relevant spin measurements are considered on the 
neutron emerging from the beam splitter BS2 by using the two spatially 
separated Stern-Gerlach devices SG1 and SG2. The two output channels of SG1 
(placed along the channel $|\psi_{3}\rangle$) are denoted S1 and S2, and s
imilarly, the two output channels of SG2 (placed along the channel 
$|\psi_{4}\rangle$) are denoted by S3 and S4.}

\end{figure}


Subsequently, a second beam splitter (BS2) is used whose reflection and 
transmission probabilities are $1/2$. After emerging from BS2, the total 
state is given by
\begin{eqnarray}
\left|\Psi\right\rangle_{BS2}= \frac{1}{2}\left[i \left|\psi_{3}\right\rangle \left(1+ e^{i\phi} \right)+ \left|\psi_{4}\right\rangle\left(1-e^{i\phi} \right)\right]\left|\chi\right\rangle
\end{eqnarray}
where $\langle\psi_{3}\left|\psi_{4}\right\rangle=0$, while the output `path' 
states $\left|\psi_{3}\right\rangle$ and $\left|\psi_{4}\right\rangle$ are 
unitarily related to the  states $\left|\psi_{1}\right\rangle$ and 
$\left|\psi_{2}\right\rangle$ by the following relations
\begin{eqnarray}
\left|\psi_{1}\right\rangle\rightarrow\frac{1}{\sqrt{2}}\left[i\left|\psi_{3}\right\rangle+\left|\psi_{4}\right\rangle\right]; \>\>
\left|\psi_{2}\right\rangle\rightarrow\frac{1}{\sqrt{2}}\left[i\left|\psi_{4}\right\rangle+  \left|\psi_{3}\right\rangle\right]
\end{eqnarray}
Finally, the relevant spin measurements are considered for the particle 
emerging from BS2 by using the Stern-Gerlach devices SG1 and SG2 placed along 
the channels $|\psi_{3}\rangle$ and $|\psi_{4}\rangle$ respectively, with the 
inhomogeneous magnetic field within these devices oriented along the 
$+\widehat{z}$-axis. This completes the description of the setup that is 
required for the information transfer scheme considered here. 

\emph{The scheme}

We now explain how this scheme works.
For sending information to Bob, the state given by Eq.(2) is the state Alice 
prepares by suitably adjusting  the parameters $\delta$ and $\phi$ 
corresponding to spin and path degrees of freedom respectively. In order to 
illustrate the present scheme, we consider two possible choices for each of 
the parameters $\delta$ and $\phi$. Thus, there are four possible combinations 
of choices, each choice corresponding to a combined unitary operation 
performed by Alice. These four possible unitary operations are denoted by 
$U1$, $U2$, $U3$ and $U4$ where $U1$ corresponds to $\delta=0, \phi=0$, 
and similarly $U2$ ($\delta=0,\phi=\pi$), $U3$ ($\delta=\pi, \phi=0$) and 
$U4$ ($\delta=\pi, \phi=\pi$).

Now, suppose Alice wants to communicate a certain outcome by subjecting the 
particle with her to a specific unitary operation, say,  $U1$. Then, from 
Eq.(3), it follows that the particle communicated to Bob ends up being in the 
channel $|\psi_{3}\rangle$ with the spin state $|\uparrow\rangle_{z}$. 
Similarly, for the respective unitary operations $U2$, $U3$ and $U4$, in any 
given case, the particle ends up in one of the two channels  
$\left|\psi_{3}\right\rangle$ and $\left|\psi_{4}\right\rangle$, with one of 
the two spin states $\left|\uparrow\right\rangle_{z}$ and 
$\left|\downarrow\right\rangle_{z}$. All these four possibilities are 
encapsulated as follows
\begin{eqnarray}
\left|\Psi^{U1}\right\rangle_{BS2}=  \left|\psi_{3}\right\rangle \left|\uparrow\right\rangle_{z}; \>\>
\left|\Psi^{U2}\right\rangle_{BS2}=  \left|\psi_{3}\right\rangle \left|\downarrow\right\rangle_{z} \\
\left|\Psi^{U3}\right\rangle_{BS2}=  \left|\psi_{4}\right\rangle \left|\uparrow\right\rangle_{z}; \>\>
\left|\Psi^{U4}\right\rangle_{BS2}= \left|\psi_{4}\right\rangle \left|\downarrow\right\rangle_{z}
\end{eqnarray} 
Thus,  in order to discern the communicated information, Bob has to perform 
spin measurements with SG1 and SG2. Let us denote the two output channels of 
SG1 (placed along the channel $|\psi_{3}\rangle$) by S1 and S2, and similarly, 
the two output channels of SG2 (placed along the channel $|\psi_{4}\rangle$) 
are denoted by S3 and S4.

It is then seen from Eqs.(5,6) that corresponding to any one of Alice's four 
combinations of unitary operations $U1, U2, U3$ or $U4$, the communicated 
particle is detected by Bob with certainty in one of the four channels S1, S2, 
S3 or  S4. This feature therefore enables our proposed scheme to be used for 
sending information about \emph{four} distinguishable outcomes via a single 
particle, essentially by using only the superposition principle \emph{without} 
requiring the use of any entangled state. Note here, that though this scheme
is accomplished with a single particle, the particle essentially carries two
`qubits' in the form of a two-level spin state and a similar ``two-level'' 
(distinguishable) superposed path state.  

Now, the question might arise as to whether it could be possible to encode
information about \emph{more than four} distinguishable outcomes using
any variant of this scheme, e.g., by creating more superpositions by using
more beam splitters. 
We will show here that if one more beam splitter is used, it is
\emph{not possible} to increase the information capacity further.

To this end let a 'path' phase shifter be applied along one of the channels,
say along $|\psi_{3}\rangle$. This operation introduces a phase
$\eta$. The beams are then reflected by mirrors $M3$ and
$M4$, respectively, and reach the third beam splitter BS3. The beam splitter
BS3 transforms the path state vectors $|\psi_{3}\rangle$ and
$|\psi_{4}\rangle$ as
\begin{eqnarray}
|\psi_{3}\rangle\rightarrow
\frac{1}{\sqrt{2}}(i|\psi_{5}\rangle+|\psi_{6}\rangle); \>\>
|\psi_{4}\rangle\rightarrow
\frac{1}{\sqrt{2}}(|\psi_{5}\rangle+i|\psi_{6}\rangle)
\label{trans1}
\end{eqnarray}
After emerging from the beam splitter BS3, the state is given by
\begin{eqnarray}
|\psi\rangle_{BS3}=\frac{1}{2\sqrt{2}}[
(-e^{i\eta}(1+e^{i\phi})+(1-e^{i\phi}))|\psi_{5}\rangle\nonumber\\
+i(e^{i\eta}(1+e^{i\phi})+(1-e^{i\phi}))|\psi_{6}\rangle]
|\chi\rangle
\label{bs3}
\end{eqnarray}
Let us suppose that the phase $\eta$ can take two different
values, say $\eta_{1}$ and $\eta_{2}$ and likewise the other phase
$\phi$ can take two different values $\phi_{1}$ and $\phi_{2}$
respectively. The four different forms of the state (\ref{bs3})
corresponding to four different values of $\eta$ and $\phi$ are
given by
\begin{eqnarray}
|\psi_{ij}\rangle_{BS3}=\frac{1}{2\sqrt{2}}[
(-e^{i\eta_{i}}(1+e^{i\phi_{j}})+(1-e^{i\phi_{j}}))|\psi_{5}\rangle\nonumber\\
+i(e^{i\eta_{i}}(1+e^{i\phi_{j}})+(1-e^{i\phi_{j}}))
|\psi_{6}\rangle]|\chi\rangle
 \label{states}
\end{eqnarray}
where $i,j = 1,2$.

In order to encode more than four distinguishable outcomes through
our scheme using the state (\ref{bs3}), 
one requires at least three pairs of the above choice of states 
(\ref{states}) to be orthogonal (Note that the parameter
$\delta$ characterizing the spin state $\chi$ can take one of two values, as
earlier). Let us assume that such a configuration
is possible for the set 
$\{|\psi_{11}\rangle, |\psi_{12}\rangle, |\psi_{21}\rangle\}$. 
Therefore, taking the
appropriate inner products of the states (\ref{states}), one finds that
the requirement of 
$\langle\psi_{11}|\psi_{12}\rangle = 0 = \langle\psi_{11}|\psi_{21}\rangle$
leads to the choice $\phi_{2}-\phi_{1}=\pi$ and $\eta_{2}-\eta_{1}=\pi$.
However, it follows that $\langle\psi_{12}|\psi_{21}\rangle \neq 0$
in this case, thus violating our assumption of more than two orthogonal states.
Further, it is possible to verify using similar arguments that no other 
combination of more than two states from the set (\ref{states}) is orthogonal.
Therefore, one is unable to encode more than four distinguishable outcomes,
or two bits of information,
by extending our scheme using more than two beam splitters.

\emph{Probabilistic dense coding}

Our above analysis pertains to a scheme of deterministic dense coding
using the spin and path variables of a single spin-$1/2$ particle whose state 
can exist in a superposition of two paths. It could be pertinent to ask
here if one could do any better by some scheme which transmits information
not exactly, but probabilistically. A variant of this question could be
to consider an initial spin state which is not polarized along a certain
direction (say) $z$, but is rather given by
\begin{eqnarray}
|\chi\rangle = \alpha|\rightarrow\rangle_x +\beta
e^{i\delta}|\leftarrow\rangle_x
\label{unpolarized}
\end{eqnarray}
and ask, how such a state would fare in dense coding.

The total path-spin state of the particle emerging from the
beam splitter BS2 is in this case given by
\begin{eqnarray}
|\psi\rangle_{BS2}=\frac{1}{2}[i|\psi_{3}\rangle(1+e^{i\phi})+|\psi_{4}\rangle(1-e^{i\phi})][\alpha|\rightarrow\rangle+\beta
e^{i\delta}|\leftarrow\rangle]
\end{eqnarray}
Corresponding to different values of the parameter $\phi$ and
$\delta$ choosing from the set $\{0,\pi\}$, the output states
which are sent to Bob are given by
(i) $|\psi\rangle_{BS2}=i|\psi_{3}\rangle\otimes(\alpha|\rightarrow\rangle+\beta
|\leftarrow\rangle)$, if $\phi=0$, $\delta=0$; 
(ii) $|\psi\rangle_{BS2}=|\psi_{4}\rangle\otimes(\alpha|\rightarrow\rangle+\beta
|\leftarrow\rangle)$, if $\phi=\pi$, $\delta=0$; 
(iii) $|\psi\rangle_{BS2}=i|\psi_{3}\rangle\otimes(\alpha|\rightarrow\rangle-
\beta |\leftarrow\rangle)$, if $\phi=0$, $\delta=\pi$; and
(iv) $|\psi\rangle_{BS2}=|\psi_{4}\rangle\otimes(\alpha|\rightarrow\rangle-\beta
|\leftarrow\rangle)$, if $\phi=\pi$, $\delta=\pi$.

It is to be noted that the states
$|\phi_{1}\rangle=\alpha|\rightarrow\rangle+\beta
|\leftarrow\rangle$ and
$|\phi_{2}\rangle=\alpha|\rightarrow\rangle-\beta
|\leftarrow\rangle$ are non-orthogonal states and hence cannot be
distinguished deterministically. However, protocols for probabilistic
dense coding \cite{pati} rely on the fact that non-orthogonal states can
be distinguished with some probability of success if they are
linearly independent. Therefore to distinguish the states
$|\phi_{1}\rangle$ and $|\phi_{2}\rangle$ probabilistically,  
these should be linearly independent. This follows from the fact that if
$\lambda_{1}|\phi_{1}\rangle+\lambda_{2}|\phi_{2}\rangle=0$
for some scalars $\lambda_{1}$ and $\lambda_{2}$, then 
$\lambda_{1}=\lambda_{2}=0$.
Thus the non-orthogonal states $|\phi_{1}\rangle$ and
$|\phi_{2}\rangle$ are linearly independent and hence they can be 
distinguished with some probability
of success.

Now the remaining task is to distinguish the states
$|\phi_{1}\rangle$ and $|\phi_{2}\rangle$ and to achieve this
goal, Bob performs a generalised measurement described by positive operator
valued measurements (POVM)s.
The corresponding POVM elements for the spin state are given by
\begin{eqnarray}
S_{1} &=&
\beta^{2}|\rightarrow\rangle\langle\rightarrow|+\alpha\beta(|\rightarrow\rangle\langle\leftarrow|
+|\leftarrow\rangle\langle\rightarrow|)+\alpha^{2}|\leftarrow\rangle\langle\leftarrow|{}\nonumber\\ S_{2} &=&
\beta^{2}|\rightarrow\rangle\langle\rightarrow|-\alpha\beta(|\rightarrow\rangle\langle\leftarrow|
+|\leftarrow\rangle\langle\rightarrow|)+\alpha^{2}|\leftarrow\rangle\langle\leftarrow|{}\nonumber\\ S_{3} &=&
(1-2\beta^{2})|\rightarrow\rangle\langle\rightarrow|+(1-2\alpha^{2})|\leftarrow\rangle\langle\leftarrow|
{}
\label{povm}
\end{eqnarray}
where $S_{1}+S_{2}+S_{3}=I$.

For illustration of this scheme, let us suppose that Bob is given the state
$|\psi\rangle_{BS2}=i|\psi_{3}\rangle\otimes(\alpha|\rightarrow\rangle+\beta
|\leftarrow\rangle)\equiv i|\psi_{3}\rangle\otimes
|\phi_{1}\rangle$. He then performs the measurement on the spin
state $|\phi_{1}\rangle$ described by
$\{S_{1},S_{2},S_{3}\}$. The POVM element $S_{2}$ is chosen in
such a way that $\langle\phi_{1}|S_{2}|\phi_{1}\rangle=0$ and this
indicates the fact that the probability of getting the result
$S_{2}$ is zero when the state $|\phi_{1}\rangle$ is given. Hence,
the measurement outcome may be either $S_{1}$ or $S_{3}$. If he gets
$S_{1}$ then the state is surely $|\phi_{1}\rangle$, but if he gets
$S_{3}$, the result is inconclusive. The success probability of
distinguishing $|\phi_{1}\rangle$ is
$1-\langle\phi_{1}|S_{3}|\phi_{1}\rangle=2(1-2\alpha^{2}\beta^{2})$.
If $\alpha=\beta=\frac{1}{\sqrt{2}}$, this reduces to the case of
deterministic dense coding described earlier,  with the success probability
being unity. 

It may be noted here that introducing additional beam splitters
does not help in encoding more information even probabilistically, since
there do not exist more than two linearly independent vectors in the
two-dimensional Hilbert space that is relevant for our spin-$1/2$ particle.
Nevertheless, one could split further the two channels $|\psi_3\rangle$ (into
say, $|\psi_5\rangle$ and $|\psi_6\rangle$), and  
$|\psi_4\rangle$ (into say, $|\psi_7\rangle$ and $|\psi_8\rangle$) by 
introducing beam-splitters in both of them, and then
use additional Stern-Gerlach measurement devices 
at the various channels.
However, such a scheme would correspond to creating more than two orthogonal
path states (or effectively qubits) per particle, and additional information 
may thereby be encoded. In the present analysis we have restricted ourselves
to consider two effective qubits (corresponding to a two-level spin state
and a two-level path state) for a single particle.

\emph{Discussion}

Note that, in the  usual dense coding scheme \cite{dens,mermin,mattle}, the 
path degrees of freedom of 
an individual spin-$1/2$ system are implicitly utilized in physically 
transporting one member of the EPR pair (that has the communicated information 
encoded in its spin state) to a receiver possessing the other member of the 
entangled pair, who then performs Bell-basis measurement to discern the 
transferred information. In the standard dense coding protocols using
entangled states, though the path variables of a particle are implicitly
involved in the very act of physically sending the particle from one location
to another, these path variables are never 
explicitly utilized in the mechanism of
encoding information. On the other hand, in the scheme we propose  here, 
while the EPR-Bohm entangled state is not required, it is in the act of 
physically sending a spin-$1/2$ particle that we use a variant of the 
Mach-Zehnder interferometer involving suitable manipulations of both the path 
and spin degrees of freedom.  

In our scheme we  exploit the ubiquituously 
present position coordinates
of the particle used for dense coding. Thus, we are able to transmit 
information about four distinguishable outcomes using both the spin and the
the suitably superposed path degrees of freedom of a single particle. 
The single spin-$1/2$ particle that we use, in effect,
carries two qubits with it (one through its spin, and another through the
two possible paths emergent from the beam splitter). Therefore, our scheme
does not violate the Holevo bound, albeit performing the task of dense
coding without quantum entanglement. The analysis of the probabilistic
variant of our dense coding scheme further reinforces the notion that the two
effective qubits carried by the position and path variables of the spin-$1/2$
particle can encode, at best, two bits of information. We conclude by 
emphasizing that our discussion of the proposed scheme, 
though presented in terms of neutral spin-$1/2$ particles (such as neutrons), 
works equally well  for photons with appropriate polarising and analysing 
devices. 

{\it Acknowledgements:} ASM and DH acknowledge grant of a project funded
by DST, India. This work was initiated during the visit of MABW who
thanks Queen's University, Belfast, for support. DH thanks the Centre for 
Science and Consciousness, Kolkata.


\begin{thebibliography}{99}

\bibitem{meas}
A. J. Leggett, Science, {\bf 307}, 871 (2005).

\bibitem{nonloc}
G. C. Ghirardi, arXiv: 0806.0647; N. D. Mermin, arXiv: 0808.1582;
G. C. Ghirardi and K. Wienand, 
arXiv: 0904.0931; G. C. Ghirardi, arXiv: 0904.0958.

\bibitem{info}
A. Zeilinger, Nature {\bf 438}, 743 (2005).

\bibitem{bec}
E. A. Cornell and C. E. Wieman, Rev. Mod. Phys. {\bf 74}, 875 (2002);
W. Ketterle, Rev. Mod. Phys. {\bf 74}, 1131 (2002).

\bibitem{superfl}
V. L. Ginzburg, Rev. Mod. Phys. {\bf 76}, 981 (2004); A. J. Leggett, 
Rev. Mod. Phys. {\bf 76}, 999 (2004).


\bibitem{qcomp}
See, for instance,
M. Nielsen and I. Chuang, "Quantum Computation and Quantum Information",
(Cambridge University Press, Cambridge (2000). 

\bibitem{teleport}
C. H. Bennett, G. Brassard, C. Crépeau, R. Jozsa, A. Peres, and W. K.
Wootters, Phys. Rev. Lett. \textbf{70}, 1895 (1993);
L. Vaidman, Phys. Rev. A \textbf{49}, 1473 (1994);
S. L. Braunstein and H. J. Kimble, Phys. Rev. A \textbf{49}, 1567 (1994).

\bibitem{bennett}
C. H. Bennett and G. Brassard, Proceedings of IEEE
International Conference on Computers,Systems and Signal processing,
pages 175-179, Bangalore, India, 1984; C. H. Bennett, G. Brassard and
N. D. Mermin, Phys. Rev. Lett. {\bf 68}, 557 (1992).

\bibitem{ekert}
A. K. Ekert,
Phys. Rev. Lett. \textbf{67}, 661 (1991).

\bibitem{dens}
C. H. Bennett and S. J. Wiesner, Phys. Rev. Lett. 69, 2881 (1992).

\bibitem{mermin}
N. D. Mermin, Phys. Rev. A 66, 132308 (2002).

\bibitem{jaegar}
G. Jaegar, {\it Quantum Information: An Overview}, (Springer, New York, 2006).

\bibitem{mattle} 
K. Mattle, H. Weinfurter, P. G. Kwiat and A. Zeilinger, Phys. Rev. Lett. 76, 
4656 (1996). 

\bibitem{braunstein}
S. L. Braunstein and H. J. Kimble, Phys. Rev. A 61, 042302 (2000).

\bibitem{li}
X. Li, Q. Pan, J. Jing, J. Zhang, C. Xie and K. Peng, Phys. Rev. Lett. 88, 
047904 (2002).

\bibitem{expt}
C. Schmid, N. Kiesel, U. K. Weber, R. Ursin, A. Zeilinger and H. Weinfurter,
New J. Phys. {\bf 11}, 033008 (2009).

\bibitem{pati}
A. K. Pati, P. Parashar and P. Agrawal, Phys. Rev. A {\bf 72}, 012329 (2005).


\end{thebibliography}
\end{document}